\documentclass[conference,10pt,letterpaper]{IEEEtran}
\usepackage{amsmath}
\usepackage{amssymb}
\usepackage{xcolor}
\usepackage{amsthm}
\usepackage{mathtools}
\usepackage{siunitx}
\usepackage{cuted}
\usepackage{graphicx}
\usepackage{subcaption}
\usepackage{epstopdf}

\usepackage{tabularx}
\usepackage{multirow}
\usepackage{booktabs}
\usepackage{diagbox}

\usepackage[boxruled,norelsize]{algorithm2e}
\makeatletter
\newcommand{\removelatexerror} {\let\@latex@error\@gobble}
\makeatother

\usepackage{tcolorbox}
\tcbuselibrary{many}

\newcommand{\transpose}{^{\text{T}}}
\newcommand{\hermitian}{^{\text{H}}}

\newcommand{\diff}{\text{d}}

\newtcbtheorem[]{alg}{Algorithm}{fonttitle=\bfseries}{alg}

\IEEEoverridecommandlockouts

\begin{document}

\title{Robustness Analysis of Networked\\ Control Systems with Aging Status}

\author{
	\IEEEauthorblockN{Bin~Han\IEEEauthorrefmark{1}, Siyu~Yuan\IEEEauthorrefmark{1}, Zhiyuan~Jiang\IEEEauthorrefmark{2}, Yao~Zhu\IEEEauthorrefmark{3}, and Hans~D.~Schotten\IEEEauthorrefmark{1}\IEEEauthorrefmark{4}}
	\IEEEauthorblockA{
		\IEEEauthorrefmark{1}Technische Universit\"at Kaiserslautern, \IEEEauthorrefmark{2}Shanghai University, \IEEEauthorrefmark{3}RWTH Aachen University, \IEEEauthorrefmark{4}DFKI
	}%
}


\maketitle

\begin{abstract}
As an emerging metric of communication systems, Age of Information (AoI) has been derived to have a critical impact in networked control systems with unreliable information links. This work sets up a novel model of outage probability in a loosely constrained control system as a function of the feedback AoI, and conducts numerical simulations to validate the model.
\end{abstract}


\IEEEpeerreviewmaketitle


\section{Introduction}\label{sec:introduction}

The concept of Age of Information (AoI) has been emerging in the past few years. Its history dates back to the beginning of 2010s, when it was first introduced in \cite{KGRK11minimizing} to describe the information freshness in a remote system. Since then, research interest has exploded in this field, noticing the unique stateful characteristic of AoI in contrast to the conventional memoryless metrics such as latency/delay \cite{Hsu2018age,KSU+2018scheduling,JZNC2019unified,HZJ+2019optimal}.


Recently, an intensive research interest has been raised on the control communication co-design (CoCoCo)~\cite{SHBF19wireless}. Since AoI can be exploited to easily describe the control/feedback signal delay in controlling systems, it has triggered a series of investigation on AoI in context of networked controlling systems. For example, \cite{KMHK19aoi} compares different AoI-related penalty functions as metric of optimal scheduling in wireless networked controlling systems with packet loss, while \cite{AVK19optimal} proposes the first AoI-based wireless scheduling policy in multi-loop networked controlling system. Additionally, AoI is compared in \cite{AVK+19age} with VoI as the scheduling performance metric in cellular networked controlling systems.

While most existing studies on AoI in CoCoCo generally take the state estimation error as metric, in this work we deepen the analysis to the system outage probability. Our analysis provides the closed-form solution of outage probability as function of system AoI, and implies its convexity features.


\section{System Model}\label{sec:model}
We consider a single-loop controlling system with a sensor, a controller and an actuator. The sensor periodically measures the current system state and transmit it in uplink (UL) to the controller, while the controller in synchronous generates an optimal control signal and transmit it in downlink (DL) to the actuator for execution. For simplification we assume the DL transmission to be reliable, while considering the UL transmission unreliable (due to lossy channel or limited computation capability of the controller).

With an unreliable UL, the status information at controller is not guaranteed to be timely, but can be outdated. Therefore we consider the AoI at time instant $t$, denoted by $\alpha(t)$, which is periodically updated w.r.t. the UL reception event:
\begin{equation}
	\alpha(t+1)=\begin{cases}
		1&\text{successful reception}\\
		\alpha(t)+1&\text{otherwise}
	\end{cases}
\end{equation}

We investigate a first-order discrete time control loop:
\begin{equation}
	\mathbf{x}(t+1)=\mathbf{A}\mathbf{x}(t)+\mathbf{B}\mathbf{u}(t)+\mathbf{w}(t),\label{equ:control_model}
\end{equation}
where $\mathbf{x}_{N\times 1}$, $\mathbf{u}_{M\times 1}$  and $\mathbf{w}_{N\times 1}$ are the state, control and noise vectors, respectively; while $\mathbf{A}_{N\times N}$ and $\mathbf{B}_{N\times M}$ are the system and input matrices, respectively.
Since the controller is not guaranteed to know $\mathbf{x}$ in real time, it estimates the current state $\mathbf{x}(t)$ according to the latest update, according to \cite{AVK19optimal}
\begin{equation}
	\begin{split}
		&\hat{\mathbf{x}}(t)\overset{\Delta}{=}\mathbb{E}\left\{\mathbf{x}(t)~\vert~\mathbf{x}(t-\alpha(t))\right\}\\
		=&\mathbf{A}^{\alpha(t)}\mathbf{x}(t-\alpha(t))+\sum\limits_{\tau=1}^{\alpha(t)}\mathbf{A}^{\tau-1}\mathbf{B}\mathbf{u}(t-\tau)\\
		=&\mathbf{x}(t)+\sum\limits_{\tau=1}^{\alpha(t)}\mathbf{A}^{\tau-1}\mathbf{w}(t-\tau).
	\end{split}	
	\label{equ:state_est}
\end{equation}

\section{Outage Probability Analysis}\label{sec:analysis}
In industrial scenarios, it is commonly required to lock the system to an ideal state $\mathbf{x}^{\text{aim}}$. It is trivial to derive that an loosely constrained controller generates the control signal
\begin{equation}
	\mathbf{u}(t)=\mathbf{B}\hermitian\left(\mathbf{BB}\hermitian\right)^{-1}\left[\mathbf{x}^\text{aim}-\mathbf{A}\hat{\mathbf{x}}(t)\right],
	\label{equ:ideal_control}
\end{equation}
hence, we have
\begin{equation}
	\begin{split}
		\mathbf{x}(t+1)\overset{(\ref{equ:control_model}, \ref{equ:ideal_control})}{=}&\mathbf{A}\mathbf{x}(t)+\mathbf{x}^\text{aim}-\mathbf{A}\hat{\textbf{x}}(t)+\mathbf{w}(t)\\
		\overset{\eqref{equ:state_est}}{=}&\sum\limits_{\tau=0}^{\alpha(t)}\mathbf{A}^{\tau}\mathbf{w}(t-\tau)+\mathbf{x}^\text{aim},
	\end{split}
	\label{equ:system_uncertaincy}
\end{equation}

When $\mathbf{x}$ varies from $\mathbf{x}^\text{aim}$ with a significant belief, a system outage is detected, and essential measures (e.g. emergency halting) shall be taken to prevent losses. In this context, here we investigate the estimated probability of system outage. More specifically, with a linear function cost of system state $\mathbf{x}$ as $G(\mathbf{x})=\mathbf{gx}$, an outage can be defined as the event that $G(\mathbf{x})\notin[G_\text{max},G_\text{min}]$
, so the estimated outage probability is
\begin{equation}
p_\text{out}=1-\text{Prob}\left\{G_\text{min}\le G({\mathbf{x}})\le G_\text{max}\right\}.
\label{equ:outage_def}
\end{equation}
For the convenience of notation let ${G}(t)=G({\mathbf{x}}(t))=\mathbf{gx}(t)$ and $G_\text{aim}=G(\mathbf{x}^\text{aim})$, we have
\begin{equation}
{G}(t)\overset{\eqref{equ:system_uncertaincy}}{=}G_\text{aim}+\mathbf{g}\sum\limits_{\tau=1}^{\alpha(t)+1}\mathbf{A}^{\tau}\mathbf{w}(t-\tau)\overset{\Delta}{=}G_\text{aim}+\mathbf{e}(t),
\end{equation}
\begin{equation}
p_\text{out}=1-\int_{G_\text{min}}^{G_\text{max}}f_{G}(g)\diff g,
\end{equation}
where $f_{{G}}(g)$ is the probability density function of ${G}$. 

To simplify analysis we consider here w.l.o.g. that $G_\text{aim}=\frac{1}{2}(G_\text{min}+G_\text{max})$, $\mathbf{w}\sim\mathcal{N}(\mathbf{0},\mathbf{\Sigma})$, and that $\mathbf{A}$ is diagonalizable. Thus, $G$ also obviously obeys a normal distribution $\mathcal{N}(G_\text{aim},\sigma_G^2)$, and therefore
\begin{equation}
	p_\text{out}=2Q\left(\frac{\Delta G}{\sqrt{\sigma_G^2}}\right),
	\label{equ:outage_q}
\end{equation}
where $\Delta G=\frac{1}{2}\left(G_\text{max}-G_\text{min}\right)$, and the Q-function $Q(y)=\frac{1}{\sqrt{2\pi}}\int_{y}^{+\infty}\exp\left(-\frac{u^2}{2}\right)\diff u$. Furthermore, to solve $\sigma_G^2$ we diagonalize $\mathbf{A}$ with 
\begin{equation}
\mathbf{\Lambda}\overset{\Delta}{=}\mathbf{P}^{-1}\mathbf{AP}=\text{diag}(\lambda_1,\lambda_2,\dots,\lambda_N).
\end{equation}
Then let $\mathbf{w}'=\mathbf{P}^{-1}\mathbf{w}$ with $\mathbf{\Sigma}'=\text{diag}(\sigma_1'^2,\sigma_2'^2,\dots,\sigma_N'^2)$ as its variance matrix, and let $\mathbf{g}'=\mathbf{gP}$, we can obtain that
\begin{equation}
\begin{split}
&\sigma_G^2
=\sum_{\tau=1}^{\alpha(t)+1}g'\mathbf{\Lambda}^\tau\mathbf{\Sigma}'\left(\mathbf{\Lambda\hermitian}\right)^\tau\mathbf{g'}\hermitian
.
\end{split}
\end{equation}

Additionally, regarding optimization issues, the convexity of system metrics is often interesting. Here we notice: i) $Q(u)$ is positive, monotonically decreasing and convex in $u\ge 0$, ii) $\frac{1}{\sqrt{v}}$ is positive, monotonically decreasing and convex in $v>0$, and iii) it always holds $\sigma_G^2>0$. Hence, it is trivial to derive that $p_\text{out}(t)$ is monotonically increasing w.r.t. $\sigma_G^2$, and it has a unique inflection point $\sigma_\text{turn}^2$. More specifically, $p_\text{out}$ is convex about $\sigma_G^2$ for $\sigma_G^2\le\sigma_\text{turn}^2$, and concave otherwise. By forcing 
\begin{equation}
	\left.\frac{\diff^2p_\text{out}(t)}{\diff\left(\sigma_G^2\right)^2}\right\vert_{\sigma_G^2=\sigma_\text{turn}^2}=0
\end{equation}
we can solve that
\begin{equation}
	\sigma_\text{turn}^2=\frac{\Delta G^2}{2}
\end{equation}

\section{Numerical Validation}\label{sec:simulations}
To validate our proposed outage probability model \eqref{equ:outage_q}, we carried out numerical simulations that describe an automated truck following other trucks in a platoon, where the model is configured as Tab.~\ref{tab:sim_setup} shows. We repeated $10~000$ times Monte-Carlo test for each of various noise specifications, and the outage probability estimated by our model matches the simulation result under all specifications with good accuracy.
\begin{table}
	\centering
	\caption{Simulation setup}
	\begin{tabular}{c|c|c|c|c}
		\toprule[2px]
		$\mathbf{x}(0)=\mathbf{x}^\text{aim}$&$\mathbf{A}$&$\mathbf{B}$&$\mathbf{\Sigma}$&$\Delta G$\\\midrule[1px]
		$[-90~0~25]\transpose$&
		$\begin{bmatrix}
		1&1&0\\0&1&0\\0&0&1
		\end{bmatrix}$
		&$[\frac{1}{2}~\frac{1}{2}~0]$&
		$\begin{bmatrix}
		0&0&0\\0&\sigma_2^2&0\\0&0&0
		\end{bmatrix}$&12.5\\
		\bottomrule[2px]
	\end{tabular}
\label{tab:sim_setup}
\end{table}

\begin{figure}[!hbtp]
	\centering
	\includegraphics[width=.9\linewidth]{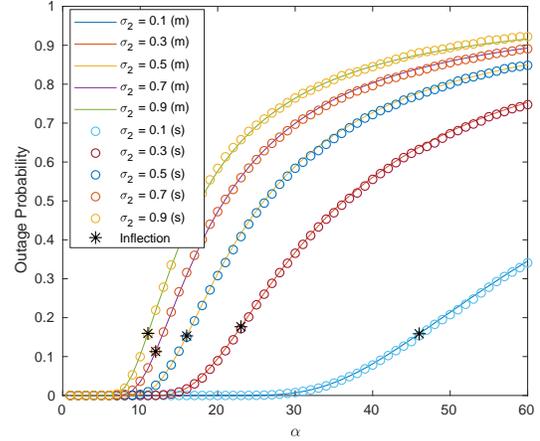}
	\caption{The simulated outage rate (``s'') match the estimation of our model (``m'') under various specifications. The estimated inflection point is labeled on every curve.}
	\label{fig:}
\end{figure}

\section{Conclusion}\label{sec:conclusion}
In this work we have investigated the robustness of networked control systems with loose constraints, proposed a novel model of system outage probability as function of the age of status information, and analyzed its convexity. The proposed model is validated by numerical simulations.


\ifCLASSOPTIONcaptionsoff
  \newpage
\fi

%
%
%
%
%




\end{document}